\documentclass[sigconf,natbib=false, nonacm]{acmart}

\usepackage{titlesec}

\titlespacing*{\section}{0pt}{5pt}{0pt} 
\titlespacing*{\subsection}{0pt}{2pt}{0pt} 
\usepackage[table]{colortbl}
\usepackage{algorithm}
\usepackage{algpseudocode}
\usepackage{enumitem}
\usepackage{graphicx} 
\usepackage{siunitx}

\RequirePackage[
    datamodel=acmdatamodel,
    style=acmnumeric,
    sorting=none
]{biblatex}
\newcommand{\ie}{\emph{i.e.}, }
\newcommand{\eg}{\emph{e.g.}, }

\copyrightyear{2025}
\acmYear{2025}
\setcopyright{cc}
\setcctype{by}
\acmConference[NOSSDAV '25]{The 35th edition of the Workshop on Network and Operating System Support for Digital Audio and Video}{March 31-April 4, 2025}{Stellenbosch, South Africa}
\acmBooktitle{The 35th edition of the Workshop on Network and Operating System Support for Digital Audio and Video (NOSSDAV '25), March 31-April 4, 2025, Stellenbosch, South Africa}\acmDOI{10.1145/3712678.3721881}
\acmISBN{979-8-4007-1469-6/2025/03}

\begin{document}

\title{End-to-End Learning-based Video Streaming Enhancement Pipeline: A Generative AI Approach}

\author{Emanuele Artioli}
\email{emanuele.artioli@aau.at}
\orcid{1234-5678-9012}
\affiliation{
  \institution{Alpen-Adria-Universitaet}
  \streetaddress{Universitaetstrasse 3}
  \city{Klagenfurt}
  \state{Kaernten}
  \country{Austria}
  \postcode{9020}
}
\author{Farzad Tashtarian}
\email{farzad.tashtarian@aau.at}
\orcid{1234-5678-9012}
\affiliation{
  \institution{Alpen-Adria-Universitaet}
  \streetaddress{Universitaetstrasse 3}
  \city{Klagenfurt}
  \state{Kaernten}
  \country{Austria}
  \postcode{9020}
}
\author{Christian Timmerer}
\email{christian.timmerer@aau.at}
\orcid{1234-5678-9012}
\affiliation{
  \institution{Alpen-Adria-Universitaet}
  \streetaddress{Universitaetstrasse 3}
  \city{Klagenfurt}
  \state{Kaernten}
  \country{Austria}
  \postcode{9020}
}


\renewcommand{\shortauthors}{Artioli et al.}

\begin{abstract}
The primary challenge of video streaming is to balance high video quality with smooth playback. 
Traditional codecs are well tuned for this trade-off, yet their inability to use context means they must encode the entire video data and transmit it to the client. 
This paper introduces ELVIS (\textbf{E}nd-to-end \textbf{L}earning-based \textbf{VI}deo \textbf{S}treaming Enhancement Pipeline), an end-to-end architecture that combines server-side encoding optimizations with client-side generative in-painting to remove and reconstruct redundant video data.
Its modular design allows ELVIS to integrate different codecs, in-painting models, and quality metrics, making it adaptable to future innovations.
Our results show that current technologies achieve improvements of up to 11 VMAF points over baseline benchmarks, though challenges remain for real-time applications due to computational demands.
ELVIS represents a foundational step toward incorporating generative AI into video streaming pipelines, enabling higher quality experiences without increased bandwidth requirements.
\end{abstract}




\thanks{The financial support of the Austrian Federal Ministry for Digital and Economic Affairs, the National Foundation for Research, Technology and Development, and the Christian Doppler Research Association, is gratefully acknowledged. Christian Doppler Laboratory ATHENA: https://athena.itec.aau.at/}

\maketitle

\section{Introduction}
\label{sec:introduction}

\begin{figure}[t]
    \centering
    \includegraphics[width=\columnwidth]{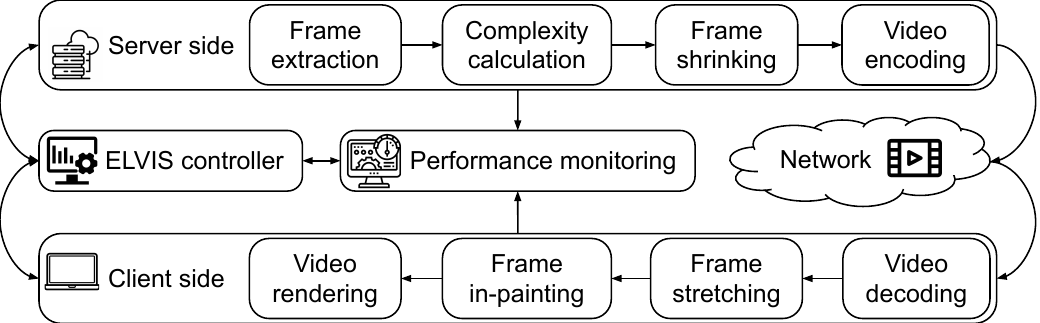}    
    \caption{Overview of the ELVIS pipeline.}\label{fig:overview}
    \vspace{-6mm}
\end{figure}

With the increasing demand for high-quality video streaming and storage, video compression methods are becoming more crucial than ever. 
Traditional codecs have significantly reduced file sizes while preserving visual quality, with each new iteration improving upon its predecessor by about 50\%~\cite{li2019codecs}. 
However, further advancements are needed to meet the growing requirements of bandwidth-constrained environments and the ever-increasing resolution of video content. 
A promising avenue is represented by neural codecs~\cite{chen2021nerv, chen2023hnerv}, \ie compressing a video into the weights of a neural network, which is then prompted by the client to recreate frames. 
In addition to server-side compression efficiency, video enhancement techniques have been explored to leverage client-side computation, such as frame interpolation and super-resolution~\cite{niklaus2017video, bao2019depth, wang2018esrgan, ledig2017photo}. 

All of the aforementioned techniques face a common limitation: they can only tackle low-level features of videos, such as edges and block-wise flow.
With the advent of large generative models, AI is now able to learn and replicate high-level video features, objects, and up to a few seconds of the whole video~\cite{openai2021dalle}. 
Therefore, a new and yet to be explored avenue for enhancing video compression is the integration of video in-painting techniques~\cite{propainter, e2fgvi, sora} that analyze the video as a whole, gather context as to what is represented in it, and fill in missing or corrupted regions.
Using the latest advances in machine learning, such as attention mechanisms~\cite{vaswani2017attention}, and training on increasingly large and curated datasets, state-of-the-art (SOTA) in-painting algorithms learn how objects typically appear and move in videos, giving them the ability to recreate far larger portions of content than previously possible~\cite{propainter, e2fgvi, sora}.

This paper's contributions are twofold: ($i$) it presents ELVIS, an innovative method that implements video in-painting alongside encoding, aiming to enhance compression efficiency by eliminating parts of the video that are challenging to encode, but can be regenerated by the client using in-painting algorithms, without significantly deteriorating the viewing experience.
This approach allows the encoder to focus on portions that cannot be easily replicated at the client side, thereby increasing video quality without additional bandwidth requirements.
The effectiveness of this method is evaluated using a variety of metrics, 
to ensure that the reconstructed video meets the high standards required for practical deployment.
Contribution ($ii$) is the release~\footnote{Codebase available at~\hyperlink{https://github.com/emanuele-artioli/elvis}{https://github.com/emanuele-artioli/elvis}}
of an end-to-end pipeline, outlined in Figure~\ref{fig:overview}, that complements video encoders on the server side, with SOTA AI techniques that remove redundancies, enhances the client side decoding with video in-painting techniques, and orchestrates and monitors the performance in terms of resources required and output quality of the whole system. This pipeline is designed to be modular and adaptable, allowing for the easy incorporation of new and improved components, such as new codecs, in-painting models, or video quality metrics. As a result, the proposed architecture not only provides immediate benefits in terms of quality improvements, but also ensures long-term adaptability as better techniques become available.


\section{Background and Related Work}
\label{sec:related_work}

Video streaming's challenges have traditionally been ($i$) increasing encoding efficiency on the server side and ($ii$) quickly adapting to network changes on the client side. 
Over time, codecs have advanced to the point where further significant reductions in video file size often require a tenfold increase in encoding time~\cite{minopoulos2020codecs-for-live}, while Adaptive Bitrate (ABR) algorithms have become adept at making efficient use of growing network bandwidths. 
With the advent of powerful AI models and mobile processors, it is now possible to share the computational burden, increasing the quality of the encoded video on the client side.

\subsection{Redundancy Removal}
Traditional codecs, such as AVC/H.264~\cite{overviewAVC}, HEVC/H.265~\cite{overviewHEVC}, and more recent iterations like VVC/H.266~\cite{overviewVVC} and AV1~\cite{overviewAV1}, have achieved significant gains in compression efficiency by minimizing spatial and temporal redundancies, \ie searching for similar groups of pixels that can be encoded once and referenced multiple times. 
Despite their success, these codecs face challenges in handling very high-resolution video content and real-time encoding due to the server-side computational complexity of their brute-force search. 

Neural codecs like NeRV~\cite{chen2021nerv} and its successor HNeRV~\cite{chen2023hnerv} have emerged as innovative alternatives by learning compact representations of video content using deep neural networks so that their weights can be sent to the client, effectively modeling redundancies as implicit features optimized for compression. 
This results in important gains in compression efficiency, but shifts the computation load to the client-side decoding, preventing their widespread adoption in practical streaming scenarios.

\subsection{Video Restoration}
The primary task on the client side is to reconstruct the encoded video with the highest quality.
To achieve this, in addition to employing an efficient decoding algorithm, various techniques can be applied to further enhance the video quality. 
Two approaches are prominent:
($i$) \textit{frame interpolation}, which focuses on generating intermediate frames between existing ones, for smoother playback, especially in low frame-rate content~\cite{niklaus2017video, bao2019depth, shi2022vfit},
($ii$) \textit{super-resolution}, \ie reconstructing high-resolution frames from low-resolution inputs by focusing on texture refinement and temporal coherence~\cite{wang2018esrgan, cao2021video, wang2024multi}.
However, the computational overhead associated with real-time applications limits their practical deployment~\cite{liu2022video}.
The two techniques have also been used jointly, for additional consistency~\cite{kim2020fisr} and, while demonstrating considerable promise, they are limited to low-level enhancement such as edge sharpening and localized motion consistency,
limiting their effectiveness in more demanding scenarios. For example, artifacts such as ghosting and unnatural textures or movements are common when dealing with highly dynamic or occluded areas~\cite{liu2022video}. These limitations can only be overcome by approaches able to leverage high-level semantic understanding.

Video in-painting algorithms, focusing on the challenge of reconstructing missing or occluded regions within video frames by exploiting spatial and temporal cues from surrounding areas, have been around for decades, initially relying on texture synthesis and patch-based methods ~\cite{criminisi2003inpainting, patwardhan2005inpainting}, but struggled with maintaining temporal consistency across frames, resulting in flickering artifacts and unrealistic reconstructions~\cite{wexler2007inpainting}.

Recent advancements in generative AI have revolutionized video in-painting by introducing models both large and efficient enough to incorporate complex spatial and temporal dependencies into their context windows~\cite{zeng2020sttn, gu2023flow, e2fgvi, propainter}, overcoming the limitations of previous techniques and setting video in-painting as a promising new avenue for restoring video streaming content from limited encodings.

\subsection{Video Quality Assessment}
Video quality is traditionally evaluated using metrics such as Mean Squared Error (MSE)~\cite{mse}, Peak Signal-to-Noise Ratio (PSNR)~\cite{psnr}, Structural Similarity Index (SSIM)~\cite{ssim} and Video Multi-method Assessment Fusion (VMAF)~\cite{vmaf}, which assess structural and perceptual fidelity. 
However, these metrics often struggle with generative AI-enhanced content due to its unique artifact patterns. 
Learned Perceptual Image Patch Similarity (LPIPS)~\cite{lpips}, designed for inpainted and generated images, complements traditional metrics by evaluating high-level perceptual quality, making it crucial for assessing AI-driven pipelines.

\section{System Design}
\label{sec:ELVIS}

The ELVIS pipeline as shown in Figure~\ref{fig:overview} consists of server-side processing, network transmission, and client-side restoration, with performance monitoring and an orchestrator overseeing all components through a series of experiments where configuration parameters are systematically optimized.

\subsection{ELVIS -- Server side} 
\textbf{Frame extraction.} 
On the server side, an input video is analyzed and, if necessary, split into segments based on scene changes, so that each segment contains coherent frames, without abrupt camera switches that would negatively affect subsequent steps. 
The video frames are then extracted from each segment using FFmpeg~\cite{ffmpeg}, then scaled to the target resolution as determined by the experiment’s configuration. This scaling step not only standardizes input dimensions but also facilitates consistent comparisons across experiments.

\textbf{Complexity calculation.} 
The next step in the pipeline is to calculate a metric that balances a block’s complexity ($i$) with its perceptual importance ($ii$). 
Blocks are ranked on their importance metric value with Algorithm~\ref{alg:get_coordinates_to_remove}, enabling the removal process to start with those that are both data-heavy to encode and less critical visually, ensuring optimal compression and leveraging the in-painter's strength.

\begin{figure}[b]
    \centering  \includegraphics[width=\columnwidth]{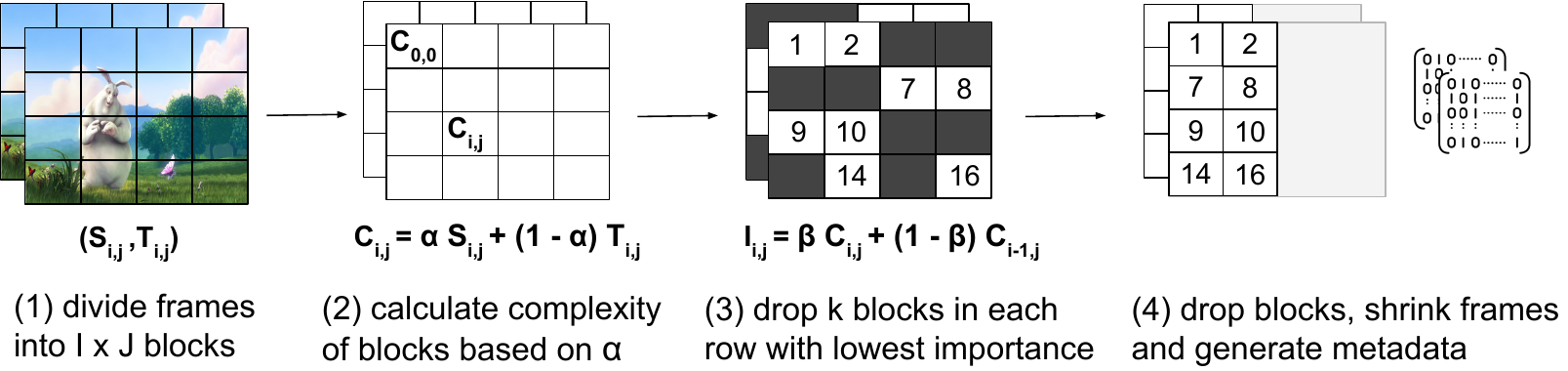}
    \caption{Complexity calculation and frame shrinking.}
    \label{fig:shrinking}
\end{figure}

\begin{algorithm}[t]
\caption{Select Blocks to Remove}
\label{alg:get_coordinates_to_remove}
\begin{algorithmic}[1]
\Require $T, S \in \mathbb{R}^{I \times J \times N}$: Temporal and Spatial complexity tensor
\Require $w,h$: Dimensions of the video
\Require $b$: Block size
\Require $\alpha \in [0,1]$: Weighting factor for spatial and temporal importance
\Require $\beta \in [0,1]$: Smoothing factor
\Require $r \in [0,1]$: Fraction of blocks to remove
\Require $M \in \mathbb{R}^{I \times J \times N}$: YOLO masks
\Ensure $R \in \mathbb{R}^{I \times J \times N}$: Coordinate tensor of blocks to remove
\State Instantiate $C \in \mathbb{R}^{I \times J \times N}$ \Comment{Importance tensor}
\For{$n = 0$ \textbf{to} $N - 1$}
    \If{$n = N - 1$}
        \State $C_{:,:,n} \gets S_{:,:,n}$
    \Else
        \State $C_{:,:,n} \gets \alpha \cdot S_{:,:,n} + (1 - \alpha) \cdot T_{:,:,n}$ \Comment{Equation~\ref{eq:i1}}
    \EndIf
\EndFor
\For{$n = 0$ \textbf{to} $N - 1$}
    \For{$i = 0$ \textbf{to} $I$}
        \For{$j = 0$ \textbf{to} $J$}
            \If{$M_{i,j,n} > 0$}
                $C_{i,j,n} *= -1$ \Comment{Invert importance}
            \EndIf
        \EndFor
    \EndFor
    \For{$i = 0$ \textbf{to} $I - 1$}
        \State Instantiate $row \in \mathbb{R}^{J}$
        \State $row \gets C_{i,:,n}$
        \If{$C_{i,:,n-1} \neq None$}
            \State $row \gets \beta \cdot row + (1 - \beta) \cdot C_{i,:,n-1}$ \Comment{Equation~\ref{eq:i2}}
        \EndIf
        \State \textbf{Sort} $row$ \textbf{descending}
        \State $row_{r\cdot J:} \gets None$ \Comment{Select r*J least important blocks}
        \State $R_{i, :, n} \gets row$ 
    \EndFor 
\EndFor
\State \Return $R$
\end{algorithmic}
\end{algorithm}
To gauge ($i$), we leverage Enhanced Video Complexity Analyzer (EVCA)~\cite{evca}, a tool that splits each video frame by block, \ie square group of pixels, whose size is indicated by the experiment configuration, and analyzes each block to produce two tensors for spatial and temporal complexity. 
The spatial complexity tensor measures the level of detail and variability within each block, in a frame, with higher values indicating intricate textures or fine details. 
Temporal complexity, on the other hand, evaluates the degree of motion or change between consecutive blocks, with higher values denoting rapid or irregular movements. 
These tensors have shape: $(I, J, N)$, where $I$ is the number of blocks per row in a frame, $J$ the number of blocks per column, and $N$ the number of frames.
The 3D tensors can be sliced along the frame dimension into two 2D matrices $S_{i,j}$ and $T_{i,j}$, as shown in Figure~\ref{fig:shrinking}, step (1-2).
The complexity of each block is determined by aggregating its spatial and temporal complexity using a weighted formula shown in Eq.~\ref{eq:i1}:
\begin{equation}
    \label{eq:i1}
    C_{i,j} = \alpha \cdot S_{i,j} + (1 - \alpha) \cdot T_{i,j}
\end{equation}
where $\alpha$ is a configuration parameter controlling the relative influence of spatial and temporal factors. 
This flexibility allows ELVIS to determine the optimal influence of spatial vs. temporal complexity on a per experiment case.
Temporal complexity for a frame depends on information from the subsequent frame, making the last frame solely reliant on spatial complexity.

To evaluate ($ii$), we consider how the background of a video is much less looked at than the objects in the foreground. Thus, we leverage You Only Look Once (YOLO)~\cite{redmon2016yolo} to segment the shape of objects in each frame, and generate masks that signal which blocks belong to the foreground, as depicted in Figure~\ref{fig:shrinking}, step (3).
The complexity values of these foreground blocks are multiplied by \num{-1}, to deprioritize their removal compared to background blocks. 
This step also ensures that, if the algorithm runs out of background to remove, it will start removing foreground blocks from the least complex and easier to in-paint.
To further refine block removal decisions, the final importance metric is complexity smoothed across frames using the formula shown in Eq.~\ref{eq:i2}:
\begin{equation} 
    \label{eq:i2}
    I_{i, j} = \beta \cdot C_{i, j} + (1 - \beta) \cdot C_{i-1, j}
\end{equation}
where $\beta$ is another tunable parameter that controls the degree of temporal smoothing. Low values promote similar removal decisions on subsequent frames, avoiding flickering on fast moving scenes, while, for $\beta$ close to \num{1}, past decisions are disregarded, ensuring that on slow-moving scenes, in-painting models see every block every few frames.

\begin{algorithm}[t]
\caption{Frame Shrinking}
\label{alg:shrink frames}
\begin{algorithmic}[1]
\Require $c$: Color channels (3)
\Require $F \in \mathbb{R}^{I \times J \times c}$: Input frame
\Require $n$: Frame number
\Require $b$: Block size
\Require $R \in \mathbb{R}^{I \times J \times N}$: Coordinate tensor of blocks to remove
\Ensure $F' \in \mathbb{R}^{I \times J-r \cdot J}$: Shrunk frame
\Ensure $O \in \mathbb{R}^{I/r \times J/r \times N}$: Binary mask tensor

\State Instantiate $L \in \mathbb{R}^{I/b \times J/b \times b \times b \times c}$
\For{$i = 0$ \textbf{to} $I/b$}
    \For{$j = 0$ \textbf{to} $J/b$}
        \State $L_{i, j} \gets F_{i \cdot b:(i+1) \cdot b, j \cdot b:(j+1) \cdot b, :}$
    \EndFor
\EndFor

\For{$i = 0$ \textbf{to} $I/b$}
    \For{$j = 0$ \textbf{to} $J/b$}
        \If{$j \notin R_{i, j, n}$}
            \State $F'_{i, j} \gets F_{i, j}$ \Comment{Fill shrunk frame with frame block}
            \State $O_{i, j, n} \gets 0$ \Comment{Signal in mask that block was filled}
        \Else
            \State $O_{i, j, n} \gets 1$ \Comment{Signal that block was not filled}
        \EndIf
    \EndFor
\EndFor
\State \Return $F'_{i, j}$
\State \Return $O_{i, j, n}$
\end{algorithmic}
\end{algorithm}

\textbf{Frame Shrinking}. 
Having determined in which order blocks should be removed to optimize the trade-off between video file size and reconstruction quality, frames can be shrunk, \ie frame resolution can be reduced by removing the blocks identified in the previous step.
Current codecs require video frames of constant, rectangular shape; therefore, blocks need to be removed in such a way that respects this. 
Two approaches are possible: one is to remove blocks independently of each other and then rearrange them so that the result forms rectangular frames, all of the same resolution. 
For example, a frame where fewer blocks were removed, could "lend" blocks to a frame that was impacted more heavily by block removal. 
However, moving blocks from one frame to another, or from one position to another in the same frame, breaks the video continuity, and thus reduces the efficiency of codecs, which then need to overcome this discontinuity with higher information count, defeating the purpose.
The second approach is to replace removed blocks with placeholders, \eg black blocks, which might seem to contain no information, and to be therefore possible to encode at virtually no information overhead, but this is not the way modern codecs are implemented.
Indeed, there is little gain in encoding efficiency in replacing an image block with a full black one, unless that black block is repeated multiple times in rapid sequence so that it can be referenced. 
ELVIS solves this problem by iteratively shrinking frames one row at a time, removing the block at highest importance, until the frame is shrunk by the amount defined in the configuration parameters.
Thus, at each iteration, each frame is one block narrower, and can be reunited in a way that maintains both the shape and the visual consistency required by codecs, as shown in Figure~\ref{fig:shrinking}, step (4).
All the information needed to shrink frames appropriately is contained in the importance matrix, therefore the shrinking process can be highly sped up via parallelization (\ie ProcessPoolExecutor from the concurrent~\cite{concurrent} Python package), to send each frame to a different thread, that will execute Algorithm~\ref{alg:shrink frames}.

\textbf{Video Encoding}.
Once the frames are shrunk, they can be encoded with any codec, a testament to ELVIS's adaptability, and at a lower bitrate than the full frames would require, based on the amount of shrinking. 
To correctly assess the best bitrate for the large set of resolutions that shrinking introduces, we leverage ARTEMIS's mega-manifest~\cite{artemis}.
Presented experiments include the most common traditional encoder, namely AVC, and the newer iteration of learned codecs, HNeRV. We consider these encoders to be a sufficient sample to showcase the potential and versatility of ELVIS, but we will be looking into testing new components in the future. 
Minimal metadata is finally generated and represented in Figure~\ref{fig:shrinking}, step (5), instructing the client on the coordinates of the removed blocks, to reverse this process and bring back the shrunk video into full resolution.
Therefore, the size of such metadata needs to be taken into account.
To minimize the impact of the metadata's size, we save it as a 2D matrix, where each row is the concatenation of all rows from a single frame, and save every frame's metadata in a single compressed numpy array in .npz~\cite{numpy} format. 
Aside from reducing the size of this file to at most \qty{5}{\percent} of the encoded video (less the higher the resolution), such compression is also very robust and fast for the client to unpack and use. 
A more precise study in metadata compression might reduce their impact even further, but is deemed out of scope for the present study.

\subsection{ELVIS -- Client Side}

\begin{algorithm}[t]
\caption{Frame Stretching}
\label{alg:stretch_frames}
\begin{algorithmic}[1]
\Require $c$: Color channels (3)
\Require $F' \in \mathbb{R}^{I \times J-r \cdot J}$: Shrunk frame
\Require $n$: Frame number
\Require $b$: Block size
\Require $O \in \mathbb{R}^{I/r \times J/r \times N}$: Binary mask tensor
\Ensure $F \in \mathbb{R}^{I \times J \times c}$: Stretched frame
\For{$i = 0$ \textbf{to} $I/b$}
    \For{$j = 0$ \textbf{to} $J/b$}
        \If{$O_{i, j, n}=0$}
            \State $F_{i, j} \gets F'_{i, j}$ \Comment{Fill stretched frame with shrunk block}
        \Else
            \State $F_{i, j} \gets 0$ \Comment{Fill stretched frame with black block}
        \EndIf
    \EndFor
\EndFor
\end{algorithmic}
\end{algorithm}

\begin{figure}[t]
    \centering
    \includegraphics[width=\columnwidth]{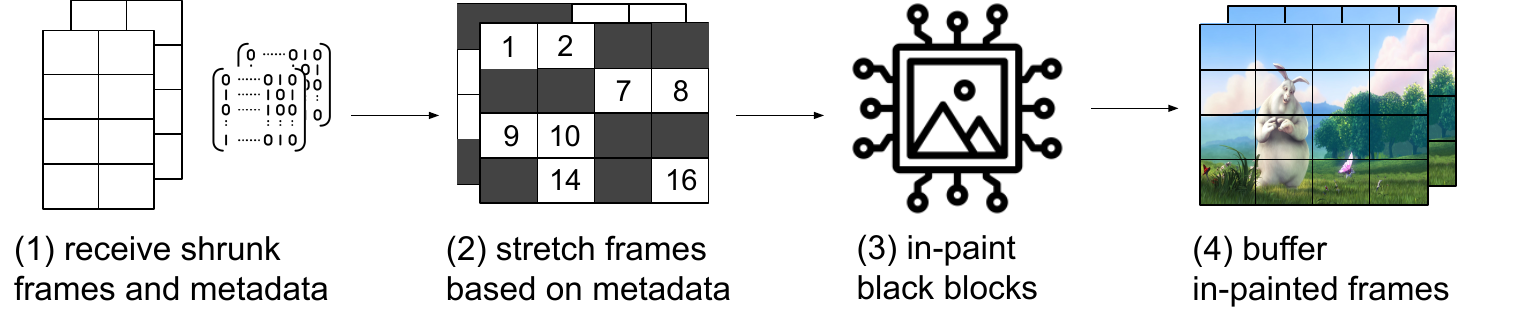}
    \caption{Frame stretching and in-painting.}
    \label{fig:stretching}
\end{figure}

At the client end, the framework decodes the received shrunk file, stretches the frames back to full resolution based on the metadata (adding temporary black blocks in place of the removed ones), and passes the video to the in-painting model to restore detail, as illustrated in Figure~\ref{fig:stretching}.

\textbf{Decoding}.
Upon receiving the encoded video and metadata, the client begins by decoding the shrunk frames. If the video was encoded using a traditional codec like AVC, FFmpeg parses the video and extracts the frames directly. For neural codecs like HNeRV, the client runs the HNeRV decoding model, which outputs padded frames. FFmpeg then removes the padding introduced during encoding, standardizing the resolution of frames regardless of the codec used. This ensures ELVIS remains agnostic to the encoding method, allowing seamless transitions into subsequent steps.

\textbf{Frame Stretching.}
Once decoded, the shrunk frames are stretched back to their original resolution by reintegrating black placeholder blocks in positions marked by the metadata, as detailed in Algorithm~\ref{alg:stretch_frames}.
For each frame, the algorithm iterates over the corresponding mask to determine whether the next block should be taken from the shrunk frame or will be in-painted. Blocks marked for in-painting are filled with black placeholders, creating stretched frames ready for reconstruction. As was the case for shrinking, the stretching process is efficiently parallelized.

\textbf{Frame In-painting.}
With the stretched frames prepared, the next step is to restore the missing content using generative in-painting models. ELVIS supports any model adhering to the standard input-output format, which includes video frames and masks as inputs and inpainted frames as outputs. For this study, two SOTA in-painting models have been tested, namely ProPainter~\cite{propainter} and E2FGVI~\cite{e2fgvi}, but the pipeline can accommodate other models with minimal configuration changes.
The in-painting model reconstructs missing blocks by leveraging spatial and temporal context from surrounding video content, plus its own understanding of the scene's context. These frames are then losslessly encoded into video using FFmpeg, completing the reconstruction process.

\section{Performance Evaluation}
\label{sec:evaluation}

\begin{table}[t]
\caption{ELVIS configuration parameters and possible instantiations.}
\vspace{-4mm}
\begin{tabular}{ll}
\hline
Parameter        & Values          \\ \hline
Video            & DAVIS dataset   \\
Block Size       & 8, 16, 32, 64          \\
Removed Fraction & 0.25, 0.50, 0.75       \\
$\alpha$         & 0, 0.25, 0.50, 0.75, 1 \\
$\beta$          & 0, 0.25, 0.50, 0.75, 1 \\
Height           & 360, 540, 720, 900     \\
Width            & 640, 960, 1280, 1600   \\
Codec            & AVC, HNeRV   \\
In-painter Tool  & ProPainter, E2FGVI
\end{tabular}
\vspace{-4mm}
\label{tab:parameters}
\end{table}

An integral part of the ELVIS architecture is monitoring the behavior and results of each component, to identify bottlenecks and improvements. Monitoring spans both server and client components, recording the time taken by each pipeline stage and evaluating the quality of the final output.

\textbf{Benchmark Comparisons.}
To assess the effectiveness of the pipeline, we run experiments in which the in-painted video is compared against a benchmark video encoded at the same resolution and file size but without client-side enhancements, across a range of metrics traditionally used for video evaluation (MSE, PSNR, SSIM, VMAF) as well as metrics specifically suited for in-painted content (LPIPS). 
This allows ELVIS to quantify the benefits of its in-painting process and determine whether the experiment's parameter configuration has resulted in meaningful improvements.

\textbf{Experiment Setup.}
We conducted a series of experiments to evaluate the performance of ELVIS, by running the pipeline on a Ubuntu 20.04 LTS server, equipped with an Intel Xeon Gold 5218 CPU with 64 cores running at \qty{2.30}{GHz}, and an NVIDIA Quadro RTX 8000 GPU having \qty{48}{GB} of memory. 
The experiments were designed to test each major component: encoding, in-painting, and quality measurement, across a varied range of parameter configurations, shown in Table~\ref{tab:parameters}. 
We processed sequences from the DAVIS dataset~\cite{davis}, known for its scene variety in terms of compositions and camera motions, and resolutions in line with video streaming requirements (480p and 1080p).
Given the computationally intensive nature of this task, we ended up running \num{200} valid experiments on \num{10} different videos, which we consider a varied dataset from which to draw valuable insights.

\subsection{Results}


\begin{figure}[b]
    \centering
    \includegraphics[width=\columnwidth]{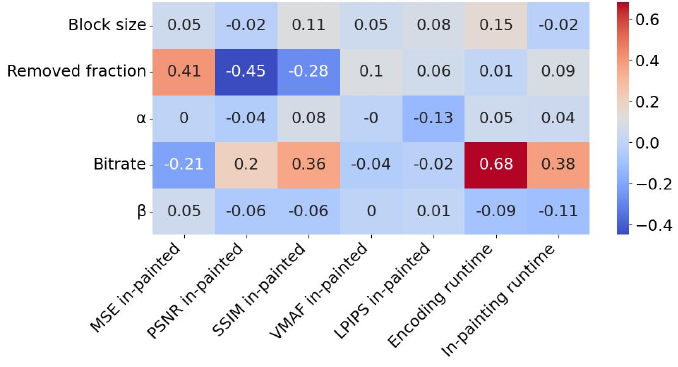}
    \caption{Impact of each parameter of ELVIS on quality metrics.}
    \label{fig:correlation}
    \vspace{-4mm}
\end{figure}

\textbf{Correlation Analysis.} Figure~\ref{fig:correlation} presents the Pearson correlation between ELVIS’s parameters and key quality metrics, including encoding and in-painting times. 
This analysis provides insights into how parameter configurations impact final video quality. 
As expected, bitrate and the fraction of blocks removed exhibit the strongest correlations. 
Higher bitrates improve video quality but increase computational demands, while larger removed fractions degrade quality by making precise in-painting more difficult. 
Interestingly, the correlation between the removed fraction and modern metrics like VMAF and LPIPS is limited, but a thorough investigation is deemed beyond the scope of this work.


\begin{figure}[b]
    \centering\includegraphics[width=\columnwidth]{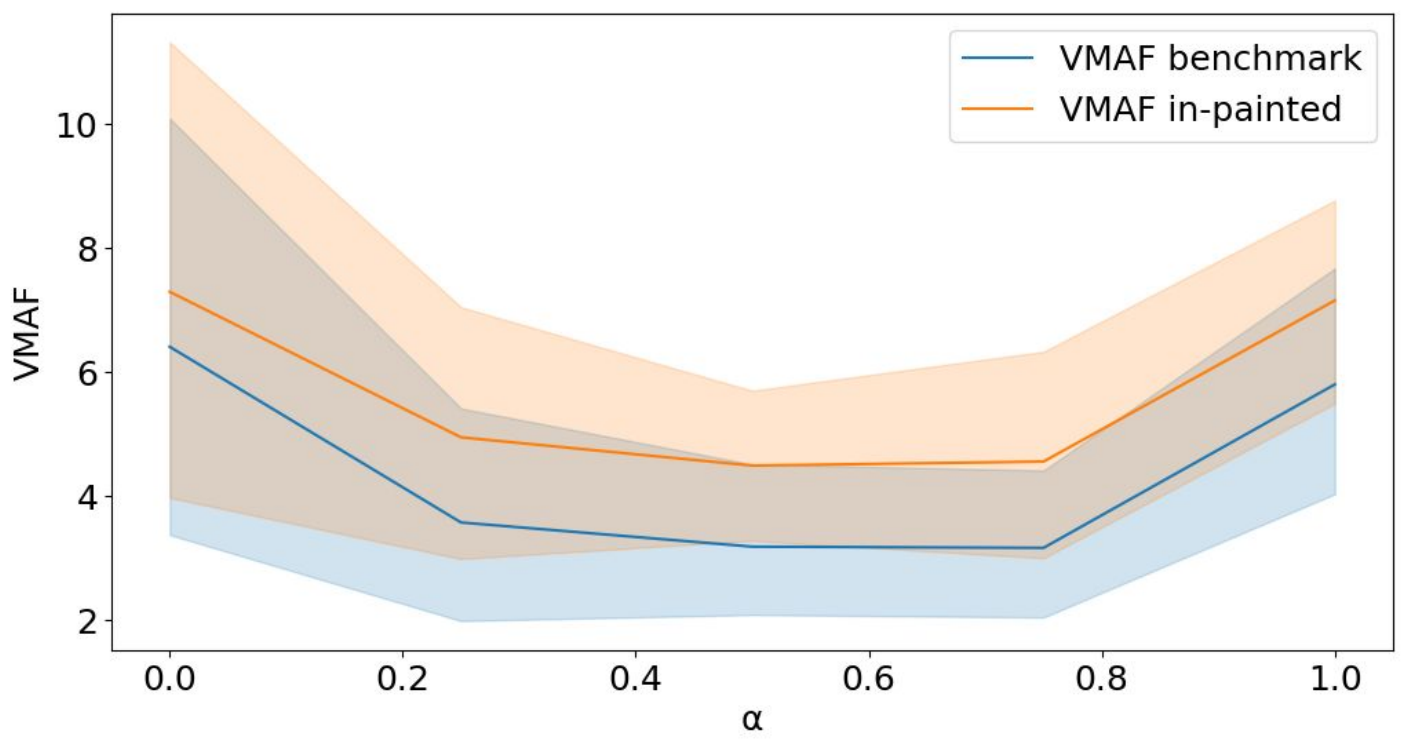}
    \vspace{-4mm}
    \caption{VMAF distribution over $\alpha$.}
    \label{fig:vmaf-alpha}
    \vspace{-4mm}
\end{figure}

\textbf{Parameter Impact.} Figure~\ref{fig:vmaf-alpha} shows how variations in $\alpha$, the spatial-to-temporal complexity weighting parameter, affect VMAF scores. 
An interesting trend is observed: values of $\alpha$ toward the extremes, which emphasize either spatial or temporal complexity in block removal decisions, tend to result in better video quality. 
Similar trends were observed for $\beta$, which controls temporal smoothing and block size.


\begin{figure*}[t]
    \centering
    \includegraphics[width=\textwidth]{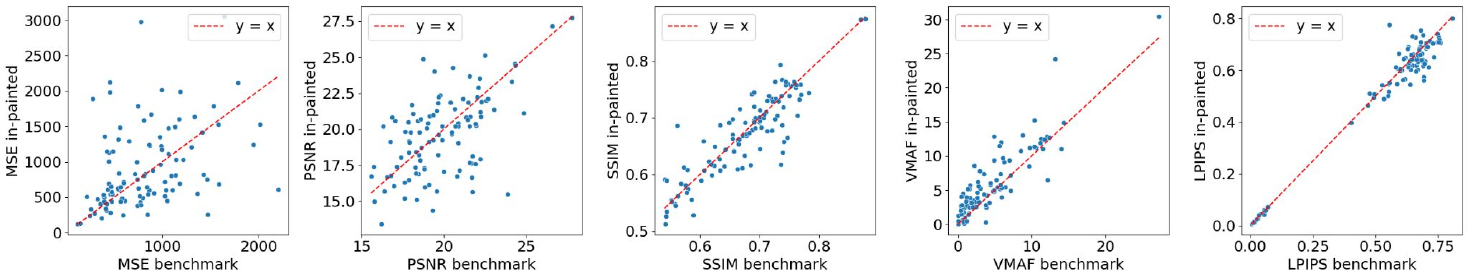}
    \caption{Comparison of benchmark and in-painted metrics.}
    \label{fig:comparison}
\end{figure*}

\textbf{Benchmark Comparison.}
Figure~\ref{fig:comparison} compares quality metrics for benchmark and in-painted videos across all experiments. 
Note that, for MSE and LPIPS, lower means better, while for PSNR, SSIM, and VMAF, higher means better.
For most cases, in-painting improves VMAF and LPIPS scores, which are the metrics most attuned to human perception, with gains of up to 10 VMAF points and 0.2 LPIPS points. 
Additionally, the overall quality tends to be relatively low, primarily due to the significant difference in resolution between the uncompressed input video and the benchmark and inpainted videos.
However, some experiments showed reduced quality, in which cases, ELVIS would transmit the benchmark video, ensuring that in-painting never degrades video quality for viewers.


\begin{figure}[b]
    \centering
    \includegraphics[width=\columnwidth]{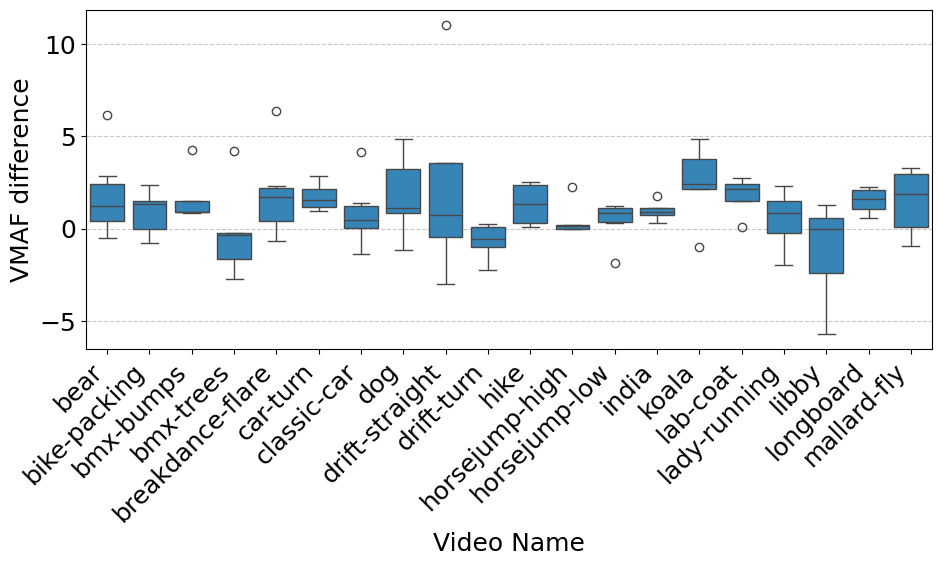}
    \caption{VMAF improvement of in-painted over benchmark, by video name.}
    \label{fig:VMAF-by-title}
    \vspace{-5mm}
\end{figure}

\textbf{Video-specific Improvements.}
Figure~\ref{fig:VMAF-by-title} highlights VMAF improvements by video title. On average, in-painting yields an increase of 2-3 VMAF points, with some videos benefiting from improvements of up to \num{11} points, corresponding to a very noticeable improvement.


\begin{figure}[b]
    \centering
    \includegraphics[width=\columnwidth]{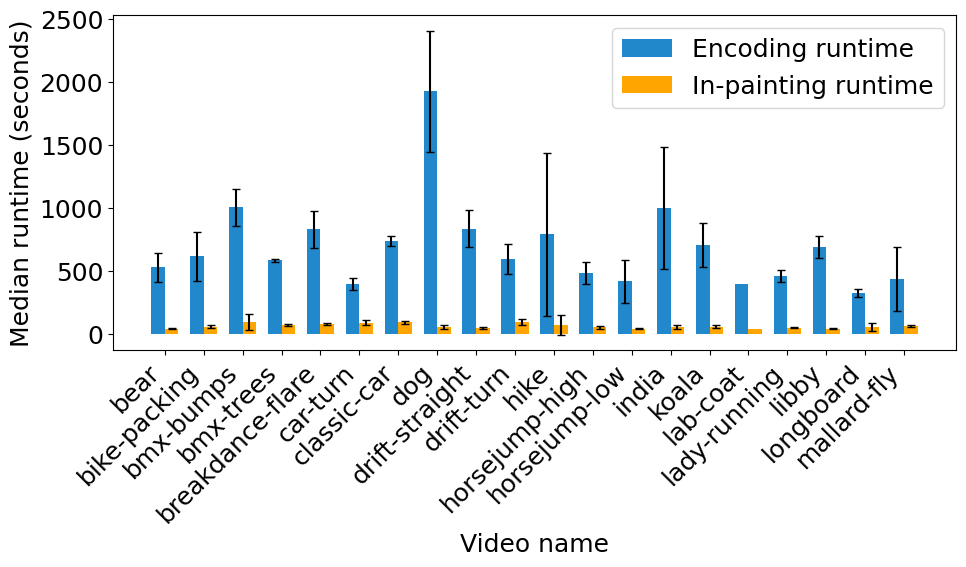}
    \caption{time complexity of encoding and in-painting, by video name.}
    \label{fig:time-complexity}
    \vspace{-5mm}
\end{figure}

\textbf{time complexity.}
Figure~\ref{fig:time-complexity} depicts the time complexity of ELVIS’s encoding and in-painting stages. 
Generative AI models, while powerful, are not yet optimized for efficiency and require significant GPU power, making higher resolution videos impossible to process.
Experiments often took \num{10} times longer than the video’s duration to complete in-painting, rendering real-time application infeasible on current consumer systems. 
Furthermore, encoding with HNeRV was an order of magnitude slower than in-painting, likely as a combination of the complexity of HNeRV and sub-optimal configuration inside of ELVIS for lack of documentation. Unfortunately, we could not reach the authors for help.
This highlights the need for advancements in both in-painting and neural encoding to support real-time streaming.


\section{Conclusions and Future Work}
\label{sec:conclusions}



This paper explored how generative AI models, unexplored in the field of video streaming, can significantly enhance video quality without increasing bitrate. 
We introduced ELVIS, an end-to-end architecture designed to repurpose general in-painting models for streaming tasks, offering a modular framework capable of integrating more powerful and efficient AI technologies as they emerge.
In a practical deployment, ELVIS would exploit server idle times, by identifying parameter configurations that optimize quality through in-painting.
Upon a client request, the server would select the best-performing configuration for delivery.
By leveraging idle server resources and ensuring fallbacks to non-inpainted videos when necessary, ELVIS guarantees that its implementation does not result in quality degradation.

While this work establishes a strong foundation, several future research directions remain to be explored. 
First, optimizing the efficiency of generative in-painting models is critical to reducing computational demands and enabling real-time applications. 
Additionally, understanding the limited correlation between certain configurations and quality metrics could lead to the introduction of different parameters and new frame shrinking strategies. 
Improving metadata compression techniques could further minimize transmission overhead, while adapting ELVIS to better support neural codecs like HNeRV, or introducing new codecs, could unlock new capabilities. 
Finally, a deeper exploration of client-side integration, including the development of lightweight in-painting models for resource-constrained devices, would significantly expand the practical applicability of ELVIS.

As the generative AI landscape evolves, ELVIS stands poised to become a key component of future video streaming systems. 
By combining advanced client-side enhancements with server-side optimizations, it offers the potential to improve user experience without increasing bandwidth consumption, marking a significant step forward in video streaming technology.


\balance
\printbibliography

\end{document}